\documentclass{article}

    \PassOptionsToPackage{numbers, compress}{natbib}

\usepackage[final]{configs/neurips_2024}

\usepackage{amsmath,amsfonts,bm}

\def\eqref#1{equation~\ref{#1}}

\def\1{\bm{1}}

\DeclareMathAlphabet{\mathsfit}{\encodingdefault}{\sfdefault}{m}{sl}
\SetMathAlphabet{\mathsfit}{bold}{\encodingdefault}{\sfdefault}{bx}{n}

\usepackage[table]{xcolor}
\definecolor{citecolor}{HTML}{0071BC}
\definecolor{linkcolor}{HTML}{ED1C24}
\usepackage[pagebackref=false, breaklinks=true, letterpaper=true, colorlinks, bookmarks=false,  citecolor=citecolor, linkcolor=linkcolor, urlcolor=gray]{hyperref}
\usepackage{orcidlink}
\usepackage{amsfonts}
\usepackage{url}
\usepackage{graphicx}
\usepackage{wrapfig}
\usepackage{tabularx}
\usepackage{color, colortbl}
\usepackage{physics}
\usepackage{booktabs}
\usepackage{multirow}
\usepackage{xcolor}
\usepackage{graphicx}
\usepackage{footnote}
\usepackage{tabularx}
\usepackage{float}
\usepackage{lipsum}
\usepackage{multicol}
\usepackage{siunitx}
\usepackage{tabularx,booktabs}
\usepackage{etoolbox}
\usepackage{algorithm}
\usepackage{listings}
\usepackage{booktabs}
\usepackage{threeparttable}
\usepackage[british, english, american]{babel}
\usepackage{subcaption}
\usepackage{enumitem}
\usepackage{adjustbox}
\usepackage{textpos}  

\usepackage{etoolbox}
\makeatletter
\AfterEndEnvironment{algorithm}{\let\@algcomment\relax}
\AtEndEnvironment{algorithm}{\kern2pt\hrule\relax\vskip3pt\@algcomment}
\let\@algcomment\relax
\newcommand\algcomment[1]{\def\@algcomment{\footnotesize#1}}
\renewcommand\fs@ruled{\def\@fs@cfont{\bfseries}\let\@fs@capt\floatc@ruled
  \def\@fs@pre{\hrule height.8pt depth0pt \kern2pt}%
  \def\@fs@post{}%
  \def\@fs@mid{\kern2pt\hrule\kern2pt}%
  \let\@fs@iftopcapt\iftrue}
\makeatother

\definecolor{deemph}{gray}{0.6}

\definecolor{LightCyan}{rgb}{0.88,1,1}
\definecolor{LightRed}{rgb}{1,0.5,0.5}
\definecolor{LightYellow}{rgb}{1,1,0.88}
\definecolor{Grey}{rgb}{0.75,0.75,0.75}
\definecolor{DarkGrey}{rgb}{0.55,0.55,0.55}
\definecolor{DarkGreen}{rgb}{0,0.65,0}

\newlength\savewidth

\newcommand{\tablestyle}[2]{\setlength{\tabcolsep}{#1}\renewcommand{\arraystretch}{#2}\centering\footnotesize}
\definecolor{baselinecolor}{gray}{.9}

\newcolumntype{x}[1]{>{\centering\arraybackslash}p{#1pt}}
\newcolumntype{y}[1]{>{\raggedright\arraybackslash}p{#1pt}}
\newcolumntype{z}[1]{>{\raggedleft\arraybackslash}p{#1pt}}

\usepackage{xspace}
\makeatletter
\DeclareRobustCommand\onedot{\futurelet\@let@token\@onedot}
\def\@onedot{\ifx\@let@token.\else.\null\fi\xspace}

\makeatother

\colorlet{darkgreen}{green!65!black}
\colorlet{darkblue}{blue!75!black}
\colorlet{darkred}{red!80!black}
\definecolor{lightblue}{HTML}{0071bc}
\definecolor{lightgreen}{HTML}{39b54a}

\renewcommand{\paragraph}[1]{\vspace{0.0em}\noindent\textbf{#1}}
\newcommand{\app}{\raise.17ex\hbox{$\scriptstyle\sim$}}

\definecolor{shadecolor}{RGB}{150,150,150}
\newcommand{\magecolor}[1]{\par\noindent\colorbox{shadecolor}}

{
\setlength{\leftmargini}{9pt}%
\begin{itemize}%
\setlength{\itemsep}{0pt}%
\setlength{\topsep}{0pt}%
\setlength{\partopsep}{0pt}%
\setlength{\parskip}{0pt}}%
{\end{itemize}}

\graphicspath{{./figures/}}

\usepackage{listings}
\usepackage{xcolor}

\lstdefinelanguage{Stata}{
	morekeywords={regress, summarize, gen, egen, drop, keep, if, foreach, forvalues, local, global, use, merge, append, sort, gsort, label, est, coefplot, graph, reghdfe, outreg2},
	sensitive=true,
	morecomment=[l]*,
	morestring=[b]"
}

\begin{document}

\title{
	\mbox{\hspace{-2em} \parbox{0.8\textwidth}{Climate Risk and Corporate Value: Evidence from Temperature Bins and Panel Regression}}
}

\author{Bo Wu$^1$\\\\
$^1$Beijing International Studies University
}

\maketitle

\begin{abstract}
\vspace{-.5em}
In empirical research, this article uses daily climate data provided by the National Oceanic and Atmospheric Administration (NOAA) of the United States to construct a temperature box with a range of 5℃, focusing on analyzing the impact of extreme high temperatures ($>$30℃) and extreme low temperatures ($\le$-10℃) on the asset value of enterprises. The results based on panel regression model show that extreme high and low temperatures can significantly reduce the asset value of enterprises. In the robustness test, this article used lagged climate data for testing, and the results still showed that extreme high temperatures had a significant negative impact on the asset value of enterprises, verifying the reliability of the benchmark regression results.

In depth heterogeneity analysis shows that in the process of addressing climate risks, companies exhibit significant differentiation based on different ownership types and industry characteristics. According to research, state-owned enterprises are relatively less affected by extreme weather due to their resource advantages and policy preferences. In addition, foreign-funded enterprises demonstrate a high level of risk resistance due to their strong management efficiency and supply chain control capabilities. In the manufacturing sector, heavy industries such as steel and communication manufacturing are particularly affected by the negative effects of extreme weather, which fully demonstrates that these industries face more severe challenges when dealing with extreme weather conditions.

\textbf{key words:  Climate Risk   Corporate Value   Temperature Bins}
\vspace{-.5em}

\end{abstract}
\section{Introduction}

\subsection{Research Background}

Previous studies have shown that extreme weather not only affects corporate profitability through direct economic losses, but also indirectly impacts corporate value through supply chain disruptions and increased financial risks. Taking the impact of Thailand's 2011 floods and rising sea levels on property prices as an example, climate risk has been widely incorporated into asset pricing by the market, and companies need to attach great importance to climate factors in investment decisions.The occurrence and severity of such events continue to escalate, posing significant threats to the normal operations and financial health of enterprises, while simultaneously presenting unprecedented challenges to corporate development and supply chain management \cite{hasna2010climate}.

This article deeply explores and clearly presents the nonlinear relationship between extreme weather and corporate value by using panel regression method with the help of daily climate data. This research result provides practical and reliable empirical support for policy makers to carry out decision-making work and business managers to carry out management practices, which can effectively assist both parties in formulating effective climate risk management strategies and promote the achievement of sustainable development goals by enterprises. These are essential to reducing carbon emissions in the economic domain—especially at the enterprise level—and to achieving national goals of peak carbon usage and carbon neutrality \cite{HARAGUCHI2015256}.

This review will focus on examining the impacts of extreme weather events at the enterprise level. By analyzing the shocks that extreme weather inflicts on businesses, and exploring strategies such as corporate strategic adjustments and proactive climate risk disclosures, this study aims to investigate effective corporate responses to such risks. Against this backdrop, the paper will specifically analyze the influence of extreme weather on corporate profitability, the responsive measures undertaken by enterprises, and the evaluation process and outcomes from the perspective of investors.
 
\subsection{Research Significance}

\paragraph{Theoretical Significance}. The innovation of this study lies in its use of the temperature bin method based on daily meteorological data to analyze the impact of climate change on firm value through high-frequency data. Compared with approaches such as the Heating Degree Day (HDD) method, this provides a more comprehensive and robust estimation. By leveraging the temperature bin technique, the study captures the nonlinear impacts of extreme weather events on business operations. Additionally, the research investigates the heterogeneous effects of climate change on enterprises across dimensions such as ownership structure and industry type, thereby exploring supply-side adaptation strategies to climate change. These findings are expected to enrich the theoretical understanding of how climate risks affect corporate operations.

\paragraph{Practical Significance}. In recent years, the frequency and intensity of climate risks induced by extreme weather have been on the rise, making the physical risks stemming from climate change a focal point of attention both domestically and internationally. In the economic realm, climate change exerts a substantial impact on corporate value, particularly on manufacturing production. Investigating the relationship between climate change and corporate production carries significant economic and social implications, providing theoretical support and policy recommendations for China’s climate adaptation strategies and the implementation of its dual-carbon goals.

\subsection{Literature Review}

\paragraph{The Economic Impact of Climate Risk on Corporate Value}. Extreme weather events have exerted significant influence on the global economy, particularly through their disruptions to supply chain networks. Haraguchi et al. (2015), in their study on the 2011 Thailand floods, pointed out that the floods had a devastating impact on Thailand’s automotive and electronics industries, thereby affecting the national economy as a whole \cite{BERNSTEIN2019253}. This event highlighted the vulnerability of enterprises in the face of extreme weather, as well as their dependence on global supply chains. Simultaneously, extreme weather events increase firms' financial risk. Bernstein et al. (2019) conducted a study on the impact of sea-level rise on real estate prices. Their findings revealed that, compared to similar properties not exposed to sea-level rise risks, those subject to such risks sold at approximately 7\% lower prices. This outcome suggests that the market has incorporated long-term climate risk into asset pricing, implying that enterprises should integrate such factors into their investment decision-making frameworks. Evidently, extreme weather can exert broad and profound impacts on corporate profitability \cite{brown2021weathering}.

Ortiz-Bobea et al. (2023) conducted an in-depth analysis of the effects of extreme temperatures on the profitability of firms across different industries, offering highly valuable insights into this field. Their empirical findings indicate that both heatwaves and cold spells significantly affect over half of all industries, with such impacts being bidirectional—some industries suffer adverse effects while others may benefit. Furthermore, the magnitude and direction of these impacts vary considerably across different seasons and industries. For instance, extreme heat and late spring cold snaps have a greater impact on corporate profitability, whereas the financial effects in autumn are relatively muted. These findings underscore the heterogeneous nature of climate risk’s influence on corporate financial performance and reflect the varying degrees of vulnerability among firms when confronted with different types of climate risks.

Moreover, physical climate risks not only affect firms’ financial costs but can also directly disrupt normal operations. Weather disasters such as cold spells and freezing events can severely affect corporate cash flow, forcing firms to rely on increased debt to hedge against shortfalls in liquidity. To manage physical climate risks, firms may optimize the use of bank credit lines and expand their credit access. This demonstrates that commercial credit and credit line management are effective tools in coping with cash flow volatility, especially for small borrowers with strong repayment capabilities (Brown et al., 2021) \cite{correa2022rising}. As awareness of climate risk deepens across various sectors, green finance and sustainable investment have emerged as prevailing trends in corporate financing. Firms must consider how to leverage green financial products to effectively reduce funding costs. At the same time, by improving their performance in Environmental, Social, and Governance (ESG) dimensions, they can bolster their resilience to weather-related disasters such as heatwaves, cold spells, and sea-level rise.

In parallel, some scholars have conducted comprehensive analyses on the relationship between extreme weather and corporate borrowing costs. Correa et al. (2022) explored the effects of climate change-induced natural disasters on corporate loan costs. They developed a detailed and specific metric system to assess borrowers’ exposure to natural disasters, distinguishing between the direct impact of disasters on firms and lenders’ updated expectations about potential future disasters. The study revealed that, following natural disasters related to climate change, even borrowers not directly impacted but with relatively high risk levels experience increased loan spreads. Furthermore, this effect becomes more pronounced during periods of heightened public concern over climate change. These findings indicate that market participants and banks have increasingly recognized the risks posed by climate change and have begun incorporating such risks into loan pricing processes \cite{amiraslani2023trust}.

\paragraph{Corporate Strategies for Addressing Climate Risk}. Supply chain disruptions represent one of the most immediate and tangible impacts of extreme weather events on corporations. The influence of such events on businesses is multidimensional, ranging from direct economic losses, to interruptions in supply chain operations, and heightened financial risk exposure. It is imperative for firms to adopt proactive measures aimed at strengthening supply chain resilience, evaluating the latent risks posed by climate policies, and leveraging green financial instruments to tackle these multifaceted challenges. Through these strategies, companies can not only shield themselves against the shocks of extreme weather events but also sustain their competitive edge amid the broader context of climate change.

A salient example is the flood event in Thailand, during which numerous industrial zones were submerged, causing significant disruptions to supply chains and subsequently affecting various industries on a global scale. This was particularly detrimental to sectors that rely heavily on sophisticated supply chain management models, such as the automotive and electronics industries. In these sectors, the floods led to production halts and logistical delays, culminating in substantial economic losses \cite{horbach2022climate}. Specifically, businesses can enhance the resilience and continuity of their supply chains through strategies such as diversifying sources of supply, establishing backup production facilities, and improving inventory management efficiency. As Haraguchi and Lall (2015) propose, designing supply chains with greater resilience can mitigate long-term economic risks \cite{flammer2021shareholder}.

In the context of corporate responses to climate risk, Corporate Social Responsibility (CSR) and performance in the bond market emerge as key areas of concern. Amiraslani et al. (2023) investigated how CSR performance influences corporate outcomes in bond markets. In their study, firms’ environmental and social performance served as proxy indicators for social capital. The analysis revealed that during the 2008–2009 financial crisis, firms with higher social capital secured more debt financing at lower bond spreads and were able to obtain longer-term debt funding. This finding suggests that when market confidence is shaken, social capital accumulated by firms functions akin to an insurance mechanism, effectively helping reduce the cost of debt financing \cite{tang2023temperature}.

Jens Horbach et al. (2024), utilizing corporate data from Germany, conducted an in-depth analysis of the complex interplay between climate change and technological innovation. Their research highlights that firms can adopt multifaceted approaches to mitigate the economic consequences of climate change. From a regulatory perspective, governments should focus on strengthening the legal and regulatory framework, laying a solid institutional foundation for green development, and providing firms with clear and consistent behavioral guidance. From the perspective of demand-side management, firms must accurately capture dynamic shifts in market demand and actively develop innovative products and services that meet consumer preferences for environmentally friendly offerings, thereby enhancing their market competitiveness. In terms of production cost management, firms can implement eco-innovative measures, such as adopting energy-efficient technologies and optimizing production processes, to reduce energy consumption and operational costs, ultimately achieving both higher productivity and improved economic performance \cite{gonzales2019goal}.

The impact of extreme climate events at the corporate level should not be underestimated. This impact is evident not only in the dimension of corporate profitability but also in prompting companies to enhance the extent of voluntary climate risk disclosure. From the perspective of investor response, there has been a positive reaction to such corporate disclosure behavior, clearly reflecting investors’ strong concern for transparency of information and the state of preparedness in companies' climate risk management processes. Relevant studies underscore the central role of climate risk management in corporate strategic planning and open up novel perspectives and directions for future academic research.

Viswanathan et al. (2021) conducted an in-depth investigation into the underlying mechanisms through which shareholder activism drives companies to increase voluntary disclosure of climate risks in the absence of mandatory disclosure regulations. Their findings indicate that environmental shareholder activism significantly promotes corporate voluntary climate risk disclosure. This effect is particularly pronounced when such activism is initiated by institutional investors, especially long-term institutional investors. Moreover, the study found that companies opting for voluntary disclosure of climate risk information following environmental shareholder activism actions attained higher market valuations post-disclosure. This phenomenon further confirms the importance investors place on information transparency regarding climate risks \cite{hatvani2016drivers}.

Additionally, Zhang Zhining et al. (2024) utilized panel data from 31 provinces in China from 2003 to 2019 to examine the causal relationships and dynamic changes between climate risk, corporate investment, and local fiscal revenue. They also explored how climate risk affects local fiscal revenues across different climatic regions based on varying levels of economic development. The study advocates that local governments should formulate fiscal and monetary policies tailored to local and temporal conditions, employing tools such as tax incentives and credit support to help climate-sensitive enterprises maintain basic operations and production continuity. Furthermore, it calls for integrated industrial park support for strategic emerging industries like low-carbon energy-saving sectors and green ecological industries, fostering the creation of a dual-carbon, green future industry and cultivating incubators for climate-adaptive new productive forces \cite{wu2024easing}.

A systematic review and integrative analysis of relevant literature reveal that extreme weather events exert profound and multi-dimensional impacts on companies. These impacts span a broad range, including corporate operational processes, dynamic cash flow management, changes in financial cost structures, and the reshaping of market trust relationships. Given the long-term and persistent nature of climate change impacts, it is essential for companies to focus on systematically building social capital to enhance resilience against shocks from extreme weather events, thereby effectively reducing the likelihood of financial risk.

Existing research has thoroughly and extensively explored the impact of extreme weather on business operations, achieving fruitful outcomes in many areas. However, significant research gaps remain in high-frequency, granular analysis of daily climate data, as well as in understanding the mechanisms by which extreme weather influences corporate asset values. Moreover, current research on corporate adaptive behavior in response to climate change lacks both breadth and depth, urgently requiring expanded and in-depth studies from a comprehensive perspective that combines macro and micro approaches. Against this backdrop, future research should aim to explore feasible strategies and implementation paths through which companies can mitigate the adverse effects of climate change by optimizing economic performance, environmental sustainability, and social responsibility. Such efforts would provide both theoretical support and practical guidance for businesses to achieve sustainable development in an increasingly volatile climate environment.

\section{Related Work}

\subsection{Data Sources}

The core variables analyzed in this paper are corporate financial data and climate data for the locations where the firms are situated. The financial data primarily concern the asset value of China’s industrial enterprises, represented by the ratio of net fixed assets to total assets. The data are sourced from the National Bureau of Statistics’ Annual Survey of Industrial Enterprises, encompassing 139,100 firms in total: 54,432 privately-owned industrial enterprises, 20,688 state-controlled industrial enterprises, 19,012 collectively-owned industrial enterprises, 32,976 mixed-ownership industrial enterprises, and 11,992 foreign-funded and Hong Kong, Macao, and Taiwan-invested industrial enterprises. The climate data include daily temperature, wind speed, and visibility for 2,843 county-level administrative divisions in China, spanning from 2005 to 2014. These data are sourced from NOAA (the National Oceanic and Atmospheric Administration of the United States).

\subsection{Baseline Model}

The primary estimation method employed in this paper is panel regression based on the industrial enterprise data and climate data, aimed at measuring the magnitude and extent of the impact of different temperature bins on the value of corporate assets. The regression equation is specified as follows.

\begin{align}
	\text{Cvalue}_{it} &= \alpha + \sum_{j=1}^{9} \beta^j \, \text{Tembin}_{it}^j + \phi X_{it} + \delta_i + \tau_t + \varepsilon_{it} 
\end{align}

$Cvalue_{it}$ denotes the ratio of net fixed assets to total assets for firm $i$ in year $t$. $Tembin_{it}^j$ represents the number of days in year $t$ during which the average daily temperature in the location of firm $i$ falls into the $j$-th temperature bin (from \textit{temp1} to \textit{temp9}, ordered from high to low temperatures). We focus primarily on the frequency of high-temperature occurrences, measuring the linear effect of different temperature bins. $X_{it}$ refers to control variables in the regression, mainly including wind speed, visibility, and atmospheric pressure at the climate level. To avoid multicollinearity, we exclude the $10^\circ\mathrm{C}$ to $15^\circ\mathrm{C}$ temperature bin from the regression and define it as the reference group.

\section{Method}

\begin{table}[t]
	\caption{
		\textbf{Descriptive Statistics}
	}
	
	\label{Descriptive Statistics}
	\vspace{.5em}
	\begin{minipage}{1.0\linewidth}{\begin{center}
				\tablestyle{6pt}{1.05}
				\begin{tabular}{lccccc}
					\hline
					\textbf{VARIABLES} & mean & standard deviation & minimum & maximum & number of observations\\
					\hline
					cvalue & 0.3559 & 0.2071 & 0.0130 & 0.9290 & 1,381,908 \\
					wind & 5.7906 & 1.5136 & 0.5888 & 15.5043 & 1,381,908 \\
					sea & 1015.75 & 1.7188 & 991.1 & 1056 & 1,381,908 \\
					visb & 8.7805 & 1.8305 & 3.5175 & 18.6411 & 1,381,908 \\
					25$\sim$30℃ & 52.6142 & 34.9701 & 0 & 189 & 1,381,908 \\
					$>$30℃ & 0.7608 & 1.8884 & 0 & 60 & 1,381,908 \\
					\hline
				\end{tabular}%
			\end{center}
	}\end{minipage}
	\vspace{-1em}
\end{table}

\subsection{Descriptive Statistics}

We next perform descriptive statistics on the data used in this paper. Table~\ref{Descriptive Statistics} shows the distribution and variability of the dependent variable cvalue, the core variable Tembin, and other control variables. Because our dataset combines a micro-level industrial enterprise survey database with daily climate data, we can capture the nonlinear influence of temperature on firms’ operations and other dynamic details. This approach provides a more comprehensive understanding of the heterogeneous effects of climate shocks and further enhances the reliability and credibility of the panel regression results.

\begin{figure}[t]
	\centering
	\hspace{-1.5em}
	\includegraphics[width=0.8\linewidth]{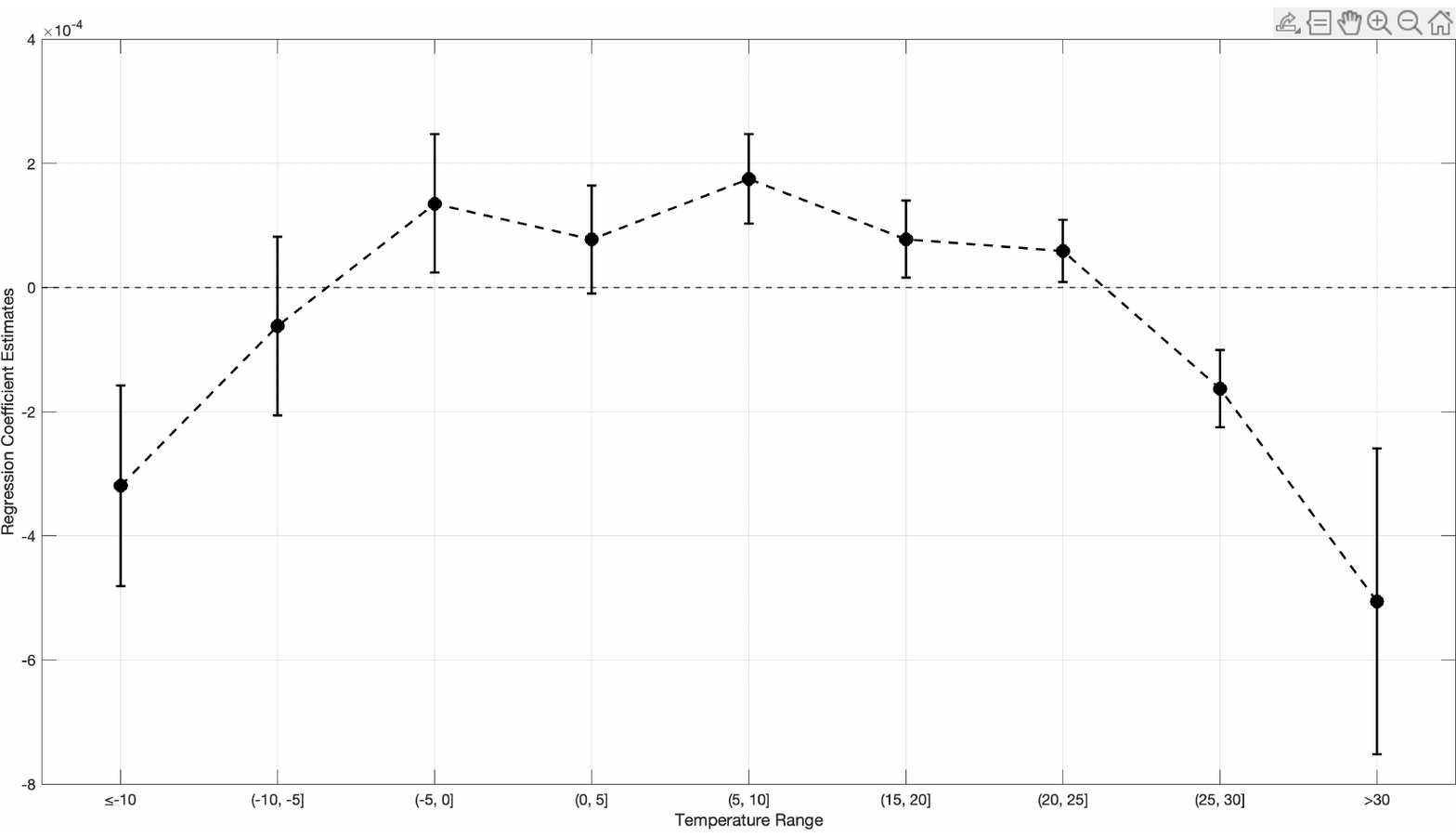}
	\caption{\textbf{Benchmark regression coefficient plot}}
	\label{Benchmark regression coefficient plot}
	\vspace{.5em}
\end{figure}


\begin{table}[t]
	\caption{
		\textbf{Benchmark Regression Results}
	}
	
	\label{Benchmark Regression Result}
	\vspace{.5em}
	\begin{minipage}{1.0\linewidth}{\begin{center}
				\tablestyle{6pt}{1.05}
				\begin{tabular}{lcccc}
					\hline
					\textbf{VARIABLES} & cvalue (1) & cvalue (2) & cvalue (3) &  cvalue (4)\\
					\hline
					$\leq -10^\circ$C  & -0.000631*** & -0.000319*** & -0.000631*** & -0.000319** \\
					& (8.10e-05)   & (8.24e-05)   & (0.000149)   & (0.000145)   \\
					-10$\sim$-5$^\circ$C & -0.000304*** & -6.22e-05    & -0.000304**  & -6.22e-05    \\
					& (7.22e-05)   & (7.35e-05)   & (0.000135)   & (0.000129)   \\
					-5$\sim$0$^\circ$C  & -0.000110**  & 0.000135**   & -0.000110    & 0.000135     \\
					& (5.57e-05)   & (5.69e-05)   & (0.000110)   & (0.000107)   \\
					0$\sim$5$^\circ$C   & -0.000112*** & 7.74e-05*    & -0.000112    & 7.74e-05     \\
					& (4.32e-05)   & (4.43e-05)   & (8.98e-05)   & (8.58e-05)   \\
					5$\sim$10$^\circ$C  & -1.12e-05    & 0.000175***  & -1.12e-05    & 0.000175**   \\
					& (3.61e-05)   & (3.67e-05)   & (7.51e-05)   & (7.27e-05)   \\
					15$\sim$20$^\circ$C & 5.15e-05     & 7.79e-05**   & 5.15e-05     & 7.79e-05     \\
					& (3.14e-05)   & (3.16e-05)   & (6.65e-05)   & (6.40e-05)   \\
					20$\sim$25$^\circ$C & 7.69e-05***  & 5.88e-05**   & 7.69e-05     & 5.88e-05     \\
					& (2.56e-05)   & (2.57e-05)   & (5.20e-05)   & (4.94e-05)   \\
					25$\sim$30$^\circ$C & -0.000192*** & -0.000163*** & -0.000192**  & -0.000163**  \\
					& (3.04e-05)   & (3.17e-05)   & (9.64e-05)   & (7.77e-05)   \\
					$>$30$^\circ$C     & -0.000190    & -0.000506*** & -0.000190    & -0.000506*   \\
					& (0.000124)   & (0.000126)   & (0.000236)   & (0.000260)   \\
					wind               &              & -0.0129***   &              & -0.0129***   \\
					&              & (0.000419)   &              & (0.00151)    \\
					sea                &              & -0.00148***  &              & -0.00148*    \\
					&              & (0.000372)   &              & (0.000823)   \\
					visb               &              & 0.00187***   &              & 0.00187      \\
					&              & (0.000407)   &              & (0.00174)    \\
					Constant           & 0.365***     & 1.904***     & 0.365***     & 1.904**      \\
					& (0.00715)    & (0.378)      & (0.0158)     & (0.837)      \\
					Controls           & N            & Y            & N            & Y            \\
					Year FE            & Y            & Y            & Y            & Y            \\
					Firm FE            & Y            & Y            & Y            & Y            \\
					Year-by-city FE    & N            & N            & Y            & Y            \\
					Observations       & 1,381,908    & 1,381,908    & 1,381,908    & 1,381,908    \\
					R-squared          & 0.142        & 0.142        & 0.142        & 0.142        \\
					\hline
				\end{tabular}%
				\caption*{\footnotesize Standard errors in parentheses. *** p$<$0.01, ** p$<$0.05, * p$<$0.1}
			\end{center}
	}\end{minipage}
	\vspace{-1em}
\end{table}

\subsection{Baseline Regression Results}

Figure~\ref{Benchmark regression coefficient plot} illustrates the baseline regression results, vividly depicting how different temperature bins affect firm value. Compared to the benchmark bin of 10°C–15°C, temperatures below –5°C or above 25°C exert a significant negative impact on corporate asset value. This finding can be attributed to the fact that when enterprises encounter extreme weather events such as cold snaps or heatwaves, workers’ health and safety are threatened, and assets and equipment depreciate more quickly, leading to negative effects on both labor and capital productivity. Consequently, firms suffer severe operational setbacks and asset value declines, manifesting in lower output and reduced productivity. As a result, in practice, firms urgently need to pay closer attention to physical climate risks in their operations.

This study further provides an in-depth analysis of the mechanisms by which physical climate risks, at different temperature intervals, affect firms’ asset values. Table~\ref{Benchmark Regression Result} presents the estimation results of the baseline panel regression model: controlling for firm fixed effects and regional macro-level variables, the first column empirically tests the effect of extreme temperature intervals on firm asset value. The results show that the regression coefficients associated with extreme high and low temperatures are both significantly negative (p $<$ 0.01), indicating that extreme temperature events notably harm firms’ asset value. To verify the robustness of this conclusion, the second column incorporates additional meteorological variables—including wind speed, sea-level pressure, and visibility—into the baseline model. The expanded model continues to find a statistically significant negative impact of extreme temperature events, with the absolute value of the coefficients slightly larger than in the baseline model. The third and fourth columns report results using city-level clustered standard errors, and for both the baseline and the expanded model, the negative impact of extreme temperature events remains significant at the 1\% level. This consistency across different statistical specifications reinforces the validity of the findings.

These results not only corroborate the theoretical hypothesis that physical climate risks exert a significant negative influence on firm asset value, but also provide micro-level empirical evidence that can inform climate risk quantification frameworks. Moreover, the findings offer practical insights for improving corporate climate risk management systems and optimizing climate-resilient investment decisions.

\begin{table}[t]
	\caption{
		\textbf{Robustness Test}
	}
	
	\label{Robustness Test}
	\vspace{.5em}
	\begin{minipage}{1.0\linewidth}{\begin{center}
				\tablestyle{6pt}{1.05}
				\begin{tabular}{lcccc}
					\hline
					\textbf{VARIABLES} & cvalue (1) & cvalue (2) & cvalue (3) &  cvalue (4)\\
					\hline
					L25$\sim$30℃ & -0.000218*** & -0.000171*** & -0.000218** & -0.000171* \\
					& (3.61e-05)   & (3.69e-05)   & (9.77e-05)  & (8.83e-05)  \\
					$>$30℃      & -0.000259**  & -0.000549*** & -0.000259   & -0.000549*** \\
					& (0.000115)   & (0.000116)   & (0.000192)  & (0.000208)  \\
					Lwind       &              & -0.0124***   &             & -0.0124***  \\
					&             & (0.000427)   &             & (0.00143)   \\
					Lsea        &              & -0.000854**  &             & -0.000854   \\
					&             & (0.000365)   &             & (0.000730)  \\
					Lvisb       &              & -0.000654    &             & -0.000654   \\
					&             & (0.000399)   &             & (0.00143)   \\
					Constant    & 0.359***     & 1.292***     & 0.359***    & 1.292*      \\
					& (0.00755)    & (0.371)      & (0.0173)    & (0.738)     \\
					Controls           & N    & Y    & N    & Y \\
					Year FE            & Y    & Y    & Y    & Y \\
					Firm FE            & Y    & Y    & Y    & Y \\
					Year-by-city FE    & N    & N    & Y    & Y \\
					Observations       & 1,358,119 & 1,358,064 & 1,358,119 & 1,358,064 \\
					R-squared          & 0.142 & 0.143 & 0.142 & 0.143 \\
					\hline
				\end{tabular}%
				\caption*{\footnotesize Standard errors in parentheses. *** p$<$0.01, ** p$<$0.05, * p$<$0.1}
			\end{center}
	}\end{minipage}
	\vspace{-1em}
\end{table}

\subsection{Robustness Test}

To further validate the baseline regression, we employ lagged climate data in the robustness test. In contemporaneous regression analyses, issues such as potential reverse causality among variables or the omission of key factors can easily lead to biased estimates. Accordingly, this study uses temperature and other relevant climate data lagged by one period, thereby reducing the likelihood of bias arising from these factors.

\begin{align}
	\text{Cvalue}_{it} &= \alpha + \sum_{j=1}^{9} \beta^j \, (L^1 \text{Tembin})_{it}^j + \phi (L^1 X)_{it} + \delta_i + \tau_t + \varepsilon_{it}
\end{align}

Table~\ref{Robustness Test} presents the regression results using the lagged data as substitutes. The analysis shows that extreme high temperatures continue to exert a statistically significant negative impact on firms’ asset values. In constructing the model, regardless of whether control variables are included or whether city-clustered fixed effects are employed, the coefficients of the core variable remain significantly negative. This finding is highly consistent with the conclusions drawn from the earlier baseline regression, providing strong evidence of the reliability of the baseline results in terms of robustness and credibility.


\begin{table}[t]
	\caption{\textbf{Ownership Heterogeneity Analysis}}
	\label{Ownership Heterogeneity Analysis}
	\vspace{.5em}
	\begin{center}
		\resizebox{\textwidth}{!}{ 
			\begin{tabular}{lccccc}
				\hline
				\textbf{VARIABLES} & Private Enterprises & State-owned Enterprises & Collective Enterprises & Mixed Enterprises & Foreign-owned Enterprises)\\
				\hline
				$\leq$-10℃  & -0.000479*** & -0.000345 & -0.000183 & -0.000465*** & 0.000153 \\
				& (0.000145)   & (0.000219) & (0.000218) & (0.000167) & (0.000355) \\
				-10$\sim$-5℃ & -0.000155    & 0.000408** & -9.47e-06  & -0.000313** & 8.75e-05 \\
				& (0.000128)   & (0.000196) & (0.000193) & (0.000152) & (0.000307) \\
				25$\sim$30℃  & -0.000224*** & -2.06e-05  & -0.000343*** & -0.000185*** & -8.39e-06 \\
				& (5.38e-05)   & (9.46e-05) & (8.84e-05) & (6.74e-05) & (0.000122) \\
				$>$30℃      & 5.79e-05     & -0.000750* & -0.00155*** & -0.000204 & -0.000164 \\
				& (0.000178)   & (0.000436) & (0.000391) & (0.000263) & (0.000586) \\
				wind        & -0.00916***  & -0.00680*** & -0.00507*** & -0.0102*** & -0.00982*** \\
				& (0.000722)   & (0.00130) & (0.00124) & (0.000909) & (0.00146) \\
				sea         & -0.00418***  & 0.00156*** & 0.000678 & -0.00220** & -0.00546* \\
				& (0.000946)   & (0.000602) & (0.000882) & (0.000854) & (0.00323) \\
				visb        & -0.00107     & -0.00152 & -0.00205* & 0.00418*** & 0.00277* \\
				& (0.000672)   & (0.00131) & (0.00113) & (0.000878) & (0.00158) \\
				Constant    & 4.659***     & -1.161* & -0.305 & 2.628*** & 5.882* \\
				& (0.959)      & (0.611) & (0.895) & (0.867) & (3.280) \\
				Controls     & Y & Y & Y & Y & Y \\
				Year FE      & Y & Y & Y & Y & Y \\
				Firm FE      & Y & Y & Y & Y & Y \\
				Observations & 535,067 & 130,288 & 167,440 & 321,108 & 109,152 \\
				R-squared    & 0.193 & 0.155 & 0.191 & 0.180 & 0.109 \\
				\hline
			\end{tabular}
		}
		\caption*{\footnotesize Standard errors in parentheses. *** p$<$0.01, ** p$<$0.05, * p$<$0.1}
	\end{center}
	\vspace{-1em}
\end{table}


\subsection{Heterogeneity Analysis}

\paragraph{Ownership Heterogeneity Analysis}. In reality, the impact of temperature shocks on firms’ asset values exhibits heterogeneity across different ownership structures and industry types. Therefore, this study conducts a heterogeneity analysis from these two perspectives. We begin with the heterogeneity arising from corporate ownership types. Following the classification standards of the National Bureau of Statistics, ownership types are generally divided into private enterprises, state-owned enterprises, collectively owned enterprises, mixed-ownership enterprises, and foreign-invested enterprises. Table~\ref{Ownership Heterogeneity Analysis} presents the heterogeneity regression results based on ownership type. The results indicate that cold snaps and other forms of extreme low temperatures pose a more pronounced negative impact on private and mixed-ownership enterprises; heat waves and other forms of extreme high temperatures impose especially prominent negative effects on collectively owned enterprises; whereas foreign-invested enterprises tend to experience comparatively lower negative impacts and physical risks under extreme weather conditions.

This evidence suggests that firms of different ownership types exhibit significant heterogeneity in their capacity to manage climate risks. Thanks to unique resource endowments and institutional advantages, state-owned enterprises suffer comparatively less from extreme weather events. This may be attributed to their close ties with the government, which enable priority access to emergency resource allocation and reduce climate risk exposure through administrative coordination. Of note, foreign-invested enterprises demonstrate marked strengths in developing climate adaptation strategies, as their asset values display significantly lower elasticity in response to extreme temperature events compared with other ownership types (p $<$ 0.01).

\paragraph{Mechanism Analysis}. The mechanism analysis indicates that the advantages of foreign-invested enterprises mainly stem from three factors. First, their international management experience leads to advanced climate risk assessment systems. Second, improvements in supply chain management efficiency foster a risk-buffering mechanism. Third, investments in technological innovation help build reserves of climate adaptation technologies. Based on the resource-based theory framework, foreign-invested enterprises integrate global knowledge networks to establish a multidimensional climate risk management system encompassing risk early warning, technological adaptation, and insurance hedging.

In terms of policy implications, this study provides micro-level evidence for climate governance under the “dual circulation” strategy. It is recommended to optimize the business environment for foreign investment through institutional innovation, with an emphasis on attracting multinational corporations possessing climate technology advantages. Such an investment attraction strategy can enhance the climate resilience of domestic enterprises via technology spillovers, thereby creating a virtuous cycle of “attracting investment–technology diffusion–capacity building.” From the perspective of industrial upgrading, enhancing climate adaptation capabilities will facilitate a transition toward low-carbon, technology-intensive production models. Consequently, this drives systematic improvements in total factor productivity and promotes the synergistic evolution of high-quality economic development.

\begin{figure}[t]
	\centering
	\hspace{-1.5em}
	\includegraphics[width=0.8\textwidth]{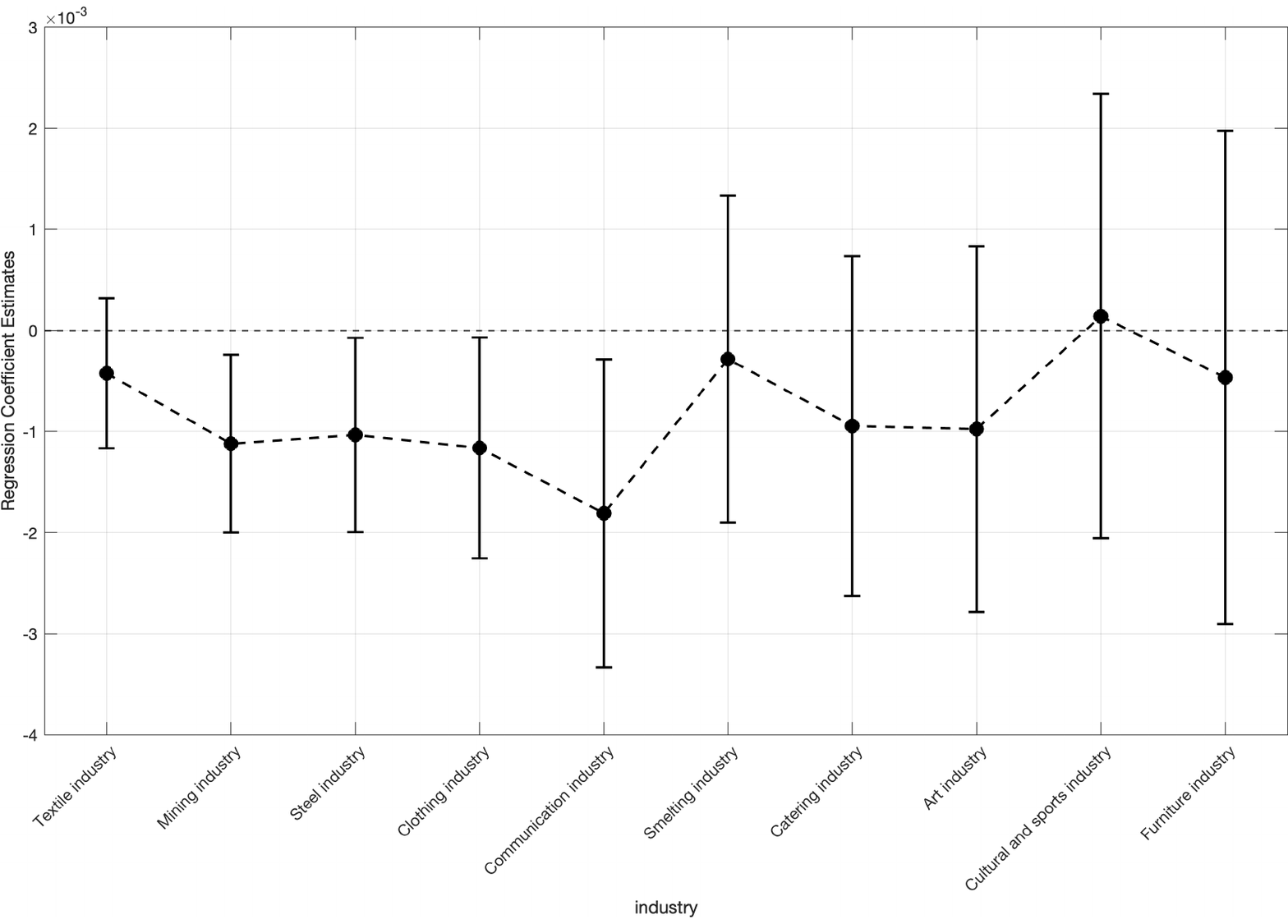}
	\caption{\textbf{Industry type heterogeneity analysis (screening model)}}
	\label{Industry type heterogeneity analysis (screening model)}
	\vspace{-1.5em}
\end{figure}

\paragraph{Heterogeneity Analysis by Industry}. Next, we conduct a heterogeneity analysis based on industry characteristics. Following the National Bureau of Statistics’ classification standards, we select representative industry types within the manufacturing sector for examination. Figure~\ref{Industry type heterogeneity analysis (screening model)} presents the results of this heterogeneity analysis. It reveals that in mining, steel, apparel, and telecommunications manufacturing, extreme high temperatures cause a significant negative impact on firms’ asset values.

On the one hand, these industries largely comprise heavy manufacturing activities that take place in open-air operating environments, rendering workers and equipment more vulnerable to direct effects of extreme weather. Cooling devices, such as air conditioners, are often less effective for climate adaptation, which magnifies the physical risks of extreme weather. On the other hand, heavy industries represented by the steel sector and ICT communications equipment manufacturing typically face high production adjustment costs and exhibit price stickiness. When extreme weather events occur, the prices of commodities and raw materials often rise. However, final product prices are constrained by existing contractual obligations and thus cannot be adjusted, reflecting price stickiness. The surge in intermediate goods costs, combined with difficulties in supply chain adjustments, makes these heavy industries particularly susceptible to climate shocks.

In contrast, industries such as culture, sports, and catering—which are primarily conducted indoors—encounter relatively more manageable physical risks. They can also adjust their prices promptly to compensate for increases in menu and “shoe-leather” costs. Accordingly, the negative impact of climate risk on the asset values of such firms is comparatively small.
\section{Implementation}

\subsection{Enhancing Corporate Resilience}

Based on the empirical analysis in this paper regarding the impact of extreme weather on corporate value, and in order to effectively respond to the economic risks brought about by climate change, enhance enterprises’ capacity to address climate risks, and promote sustainable development, the following policy recommendations are proposed:

First, the government should fully acknowledge the central role enterprises play in climate adaptation and governance, and local authorities must recognize the critical part enterprises perform in responding to climate risks. Climate adaptation and governance measures are effective tools for strengthening corporate environmental awareness and incentivizing enterprises to assume environmental responsibilities. They also represent the necessary path for adapting to extreme weather and hedging against physical climate risks. On the basis of scientifically measuring the characteristics of climate risks at the enterprise level, as well as their transmission mechanisms and loss functions, local governments should consider regional and industrial heterogeneity when making climate-related decisions \cite{XDCH202405010}.

In light of this, relevant agencies should, based on the conditions of different regions, capitalize on local circumstances to motivate enterprises to fully leverage their resource endowments, human capital, and technological innovation capabilities. Each grassroots government should optimize the business environment through institutional innovation, establish dedicated subsidies for the development of climate-resilient infrastructure and an ESG reputation evaluation system, and guide enterprises in green technological innovation. Regulatory authorities should build a climate risk stress-testing system and a green capital allocation mechanism, integrating carbon footprint certification and digital technologies to enhance the effectiveness of climate governance. Digital transformation can improve risk prevention and control through digital twin early-warning systems, blockchain traceability platforms, and AI-based energy management.

In addition, regulators should establish a climate-adaptive investment decision matrix, pilot a climate risk reserve system, and set up a national climate technology innovation fund, forming a governance framework led by the government, driven by the market, and enabled by technology. This will achieve the triple objectives of improving resilience at the micro level, transforming industries at the meso level, and realizing high-quality development at the macro level.

\subsection{Establishing Climate Adaptation Mechanisms}

Second, the government must establish systems and mechanisms at the societal level to address climate change. On the one hand, industries and enterprises of all types should actively respond to the government’s call by adopting energy-saving technologies and green management practices, making effective use of digital technologies for intelligent transformation, and fulfilling local government requirements concerning the “dual control” of energy consumption and carbon emissions. In this way, they contribute their own “green” efforts toward achieving the dual-carbon goals, working on the supply side to mitigate climate change, slow temperature rise, and reduce the occurrence of various heatwave and extreme high-temperature events.

On the other hand, enterprises should employ scientific methods to adapt to climate change from multiple dimensions—such as supply chain configuration, technology development, and human capital cultivation. As the main driving force in climate change adaptation and a key driver for social and economic operations and high-quality development, enterprises must make multifaceted, comprehensive efforts in areas including green credit funding, climate-adaptive talent development, resilient infrastructure construction, and supply continuity. By combining multiple tools to confront increasingly frequent physical climate risks and extreme weather events, enterprises can plan more effectively for possible secondary disasters—such as meteorological hazards and power shortages—under extreme weather scenarios, thus mitigating and smoothing out the negative impacts on productivity and enterprise value caused by temperature shocks \cite{CSFY202411014}.

Meanwhile, the government should provide necessary tax benefits and financial incentives for businesses that actively engage in climate adaptation measures. This support helps enterprises upgrade their equipment and enhance the resilience and reliability of various production facilities and environments, and it motivates enterprises, industry associations, and other social forces to take the initiative in participating in climate adaptation actions.

\subsection{Government Market Incentives}

Third, the government should employ various market-based incentive tools to strengthen enterprises’ risk resistance. For example, the government can provide guidance and oversight in improving the market for water resources, thereby addressing drought crises and power outages triggered by extreme heat events. This allows the entire region to enhance the accessibility and utilization of water resources for enterprises, thereby reinforcing their capacity—along with other market actors—to adapt to climate change and improving their supply chain resilience. Under climate-related physical risks, enterprises can thus update and localize their disaster preparedness and protective capabilities in a timely manner.

In this regard, carbon emission trading and the issuance of energy performance certificates serve as effective means to achieve both climate adaptation and the dual-carbon goals \cite{HJBU2024Z2015}. By incorporating the shadow price of energy-use improvements into market valuations, such mechanisms help reduce the financing pressure and borrowing difficulties enterprises face when cutting greenhouse gas emissions. They also provide support for proactively adapting to climate change and renovating production environments. Specifically, local governments need to adopt more precise and effective fiscal incentives, along with moderately accommodative credit policies, to encourage enterprises to actively respond to climate change, thereby effectively managing both physical climate risks and transition risks \cite{BLDS202303003}.

More concretely, local governments and social financing platforms should utilize multiple strategies—for instance, implementing a comprehensive and reasonable heat subsidy mechanism, establishing systematic aid programs, and introducing targeted green credit support projects—to help various enterprises effectively address and gradually adapt to climate risks. In particular, strategic emerging industries should be accorded sufficient funding and essential resources. By driving structural adjustments across industries and making full use of digital empowerment, the government can secure energy supplies while achieving climate governance objectives.

Within the policy framework of operational subsidies and monetary easing to promote climate adaptation, enterprises can employ sound tax planning and other measures to boost their development resilience and maintain supply chain stability. In this way, enterprises will demonstrate greater vitality and autonomy in climate adaptation efforts, which, in turn, significantly bolsters the broader society’s expectations and confidence in a low-carbon economy.

{\small
\bibliographystyle{configs/bib}
\bibliography{main}
}

\clearpage
\appendix
\section{Stata Code }


\begin{lstlisting}[language=Stata, label={lst:stataexample}]
	
	* Table 1: Descriptive Statistics
	summarize cvalue wind sea visb temp1 temp2 ind city_code
	
	* Figure 1: Coefficient plot of temp1 to temp9 in model a2
	reghdfe cvalue temp1 temp2 temp3 temp4 temp5 temp6 temp7 temp8 temp9 wind sea visb , absorb(i.year i.city_code)
	est store a2
	
	coefplot a2, /// 
	keep(temp1 temp2 temp3 temp4 temp5 temp6 temp7 temp8 temp9) /// 
	drop(_cons) /// 
	order(temp9 temp8 temp7 temp6 temp5 temp4 temp3 temp2 temp1) /// 
	vertical /// 
	yline(0) ciopts(recast(rcap)) scheme(s1mono)
	graph export Baseline_Regression_Coefficients.png, width(1000) replace
	
	**** Regression Analysis ****
	
	* ---------------------------
	* Table 2: Baseline Regressions
	* ---------------------------
	
	* Define base variables and optional additional variables
	local base_vars "temp1 temp2 temp3 temp4 temp5 temp6 temp7 temp8 temp9"
	local additional_vars "wind sea visb"
	
	* Initialize model counter
	local model_num = 1
	
	* Loop over inclusion of additional variables (0 = no, 1 = yes)
	foreach include_additional in 0 1 {
		
		* Loop over use of clustered robust standard errors (0 = no, 1 = yes)
		foreach include_cluster in 0 1 {
			
			* Construct model name like a1, a2, a3, a4
			local store = "a`model_num'"
			
			* Initialize variable list with base variables
			local vars "`base_vars'"
			
			* Add additional variables if required
			if `include_additional' == 1 {
				local vars "`vars' `additional_vars'"
			}
			
			* Initialize vce option
			local vce_opt ""
			
			* Add cluster option if required
			if `include_cluster' == 1 {
				local vce_opt "vce(cluster city_code)"
			}
			
			* Run regression
			reghdfe cvalue `vars', absorb(i.year i.city_code) `vce_opt'
			
			* Store regression results
			est store `store'
			
			* Increment model counter
			local ++model_num
		}
	}
	* Output all stored models to a Word document using outreg2
	outreg2 [a*] using "1.doc", replace word
	
	* ---------------------------
	* Table 3: Robustness Checks
	* ---------------------------
	
	* Keep data from 2006 onwards
	keep if year >= 2006
	
	* Define lagged base variables and optional lagged additional variables
	local base_vars_b "L1temp1 L1temp2 L1temp3 L1temp4 L1temp5 L1temp6 L1temp7 L1temp8 L1temp9"
	local additional_vars_b "L1wind L1sea L1visb"
	
	* Initialize model counter
	local model_num_b = 1
	
	* Loop over inclusion of additional variables (0 = no, 1 = yes)
	foreach include_additional in 0 1 {
		
		* Loop over use of clustered robust standard errors (0 = no, 1 = yes)
		foreach include_cluster in 0 1 {
			
			* Construct model name like b1, b2, b3, b4
			local store = "b`model_num_b'"
			
			* Initialize variable list with base variables
			local vars "`base_vars_b'"
			
			* Add additional variables if required
			if `include_additional' == 1 {
				local vars "`vars' `additional_vars_b'"
			}
			
			* Initialize vce option
			local vce_opt ""
			
			* Add cluster option if required
			if `include_cluster' == 1 {
				local vce_opt "vce(cluster city_code)"
			}
			
			* Run regression
			reghdfe cvalue `vars', absorb(i.year i.city_code) `vce_opt'
			
			* Store regression results
			est store `store'
			
			* Increment model counter
			local ++model_num_b
		}
	}
	
	* Output all stored robustness check models to a Word document
	outreg2 [b*] using "2.doc", replace word
	
	* ---------------------------
	* Table 4: Ownership Heterogeneity Analysis
	* ---------------------------
	
	* Define ownership types and corresponding conditions
	local ownership_types "private soe collective mix foreign"
	local ownership_conditions "private==1 soe==1 collective==1 mix==1 foreign==1"
	
	* Define regression variables
	local vars_c "temp1 temp2 temp3 temp4 temp5 temp6 temp7 temp8 temp9 wind sea visb"
	
	* Initialize model counter
	local model_num_c = 1
	
	* Loop over ownership types
	foreach type in `ownership_types' {
		
		* Get corresponding condition
		local condition : word `model_num_c' of `ownership_conditions'
		
		* Construct model name like c1, c2, c3, c4, c5
		local store = "c`model_num_c'"
		
		* Run regression with ownership condition
		reghdfe cvalue `vars_c' if `condition', absorb(i.year i.city_code)
		
		* Store regression results
		est store `store'
		
		* Increment model counter
		local ++model_num_c
	}
	
	* Output all stored ownership heterogeneity models to a Word document
	outreg2 [c*] using "3.doc", replace word
	
	**** Industry Heterogeneity ****
	
	* Sort and filter data
	sort ind
	gen temp_flag=1
	egen temp_sum=sum(temp_flag),by(ind)
	drop if temp_sum<10000
	
	* Sort and label data
	gsort -temp_sum -ind
	label variable temp_sum "number of firms"
	gen temp_neg=-temp_sum
	egen temp_group=group(temp_neg)
	gen hy_2digit_lx11=temp_group
	
	* Prepare for regression analysis
	label variable temp1 ""
	forvalues i=1/29{
		gen hy_`i'= temp1
		reghdfe cvalue hy_`i' wind sea visb if hy_2digit_lx11==`i' ,absorb(i.year i.city_code)
		est store m`i'
	}
	
	* Coefficient plot before filtering (including all m1 to m29)
	
	* Create local macro with all model names
	local models
	forvalues i = 1/29 {
		local models `models' m`i'
	}
	
	* Create local macro with all industry names
	local industries
	forvalues i = 1/29 {
		local industries `industries' hy_`i'
	}
	
	* Plot all regression coefficients
	coefplot (`models'), /// 
	keep(`industries') /// 
	levels(95) /// 
	vertical /// 
	yline(0) /// 
	ciopts(recast(rcap)) /// 
	scheme(s1mono)
	graph export Industry_Heterogeneity_Analysis_All.png, width(1000) replace
	
	* Coefficient plot 2 (selected industries only)
	coefplot (m1 m2 m7 m8 m12 m14 m18 m23 m26 m28, /// 
	keep(hy_1 hy_2 hy_7 hy_8 hy_12 hy_14 hy_18 hy_23 hy_26 hy_28) /// 
	levels(95)), /// 
	vertical /// 
	yline(0) /// 
	ciopts(recast(rcap)) /// 
	scheme(s1mono)
	graph export Industry_Heterogeneity_Analysis_Selected.png, width(1000) replace
\end{lstlisting}
\vspace{-.5em}


\end{document}